\pdfoutput=1
\documentclass{PoS}
\usepackage{amssymb,amsmath,amsfonts} 
\newcommand{\comma}{\;,}

\newcommand{\tr}{\mathrm{tr}}
\newcommand{\Tr}{\mathrm{Tr}}

\newcommand{\R}{\textrm{R}}
\newcommand{\AS}{\textrm{AS}}
\newcommand{\Adj}{\textrm{Adj}}

\newcommand{\Sym}{\textrm{S}}

\title{Orientifold Planar Equivalence:\\The Chiral Condensate }
\ShortTitle{Orientifold Planar Equivalence: The Chiral Condensate }

\author{Adi Armoni, \speaker{Biagio Lucini}, Agostino Patella\\
        Physics Department, Swansea University, Singleton Park, Swansea, SA2 8PP, UK\\
        E-mail: \email{a.armoni@swansea.ac.uk, b.lucini@swansea.ac.uk, a.patella@swansea.ac.uk.}}

\author{Claudio Pica\\
        Physics Department, Brookhaven National Laboratory, Upton, NY 11973, USA\\
        E-mail: \email{pica@quark.phy.bnl.gov.}}

\abstract{
The recently introduced orientifold planar equivalence is a promising tool for solving non-perturbative problems in QCD. One of the predictions of orientifold planar equivalence is that the chiral condensates of a theory with $N_f$ flavours of Dirac fermions in the symmetric (or antisymmetric) representation and $N_f$ flavours of Majorana fermions in the adjoint representation have the same large $N$ value for any value of the mass of the (degenerate) fermions. Assuming the invariance of the theory under charge conjugation, we prove this statement on the lattice for staggered quenched condensates in SU($N$) Yang-Mills in the large $N$ limit. Then, we compute numerically those quenched condensates for $N$ up to 8. After separating the even from the odd corrections in $1/N$, we are able to show that our data support the equivalence; however, unlike other quenched observables, subleading terms in $1/N$ are needed for describing the data for the symmetric and antisymmetric representation at $N$=3. Possible lessons for the unquenched case are discussed. 
}

\FullConference{The XXVI International Symposium on Lattice Field Theory \\
		 July 14 - 19, 2008\\
		 Williamsburg, Virginia, USA}

\begin{document}
\section{Introduction}
Orientifold planar equivalence~\cite{Armoni:2003gp} establishes the equivalence between observables in the bosonic, charge conjugation-invariant sector of a SU($N$) theory with $N_f$ flavours of Dirac fermions of mass $m$ in the symmetric or antisymmetric representation of the gauge group and with $N_f$ flavours of Majorana fermions of mass $m$ in the adjoint representation, in the limit for the number of colours $N$ going to infinity. For $N_f = 1$, at $N = 3$ the former is one-flavour QCD, while the latter for $m=0$ is ${\cal N} = 1$ Super Yang-Mills. Hence, orientifold planar equivalence allows us to connect exact results from SUSY to one-flavour~\cite{Armoni:2003fb} and multi-flavour (see~\cite{Armoni:2005wt} for an example concerning the multi-flavour case) QCD observables . In principle this equivalence would be a powerful tool for lattice gauge theorists, since it would enable us to model numerical data. However, the proof of the equivalence relies on the assumption that charge conjugation be a symmetry of the system~\cite{Unsal:2006pj}. While this is reckoned not to be a problem in practice~\cite{Armoni:2007rf}, a rigorous proof of the equivalence requires {\em ab initio} calculations. In this contribution, we will show the first results in that direction, by proving the equivalence of the quenched condensates in the antisymmetric (or the symmetric) representation and in the adjoint representation for any mass of the $N_f$ flavours of degenerate quarks\footnote{Unlike the case of fundamental fermions, for fermions in a two-index representation the quenched large $N$ limit does not coincide with the large $N$ limit in the full theory.}. The results discussed in this work have already appeared in~\cite{Armoni:2008nq}, to which we refer for further details. A test of the predictions of orientifold planar equivalence for meson masses at $N=3$ and $N_f=1$ has been reported in~\cite{Farchioni}.
\section{Analytic proof on the lattice}
Under the assumption of invariance of the theory under charge conjugation, orientifold planar equivalence on the lattice has been proved in~\cite{Patella:2005vx}. Making use of the same assumption, here we provide a more direct proof in the special case of the quenched quark condensate.\\
To be concrete, we take one taste of staggered fermions. For the gauge action, we choose the Wilson action for a SU($N$) lattice gauge theory:
\begin{eqnarray}
S = \beta \sum_{x,\mu > \nu} \left( 1 - \frac{1}{N}{\cal R}\mathrm{e}\,\tr{U_{\mu \nu}(x)}\right) \ , \qquad \beta = 2N/g^2 \ .
\end{eqnarray}
The partition function is 
\begin{eqnarray}
\label{zeta}
Z = \int \left( {\cal D} U \right) e ^{ - S} 
\end{eqnarray}
and vacuum expectation values of the operators are taken on the Yang-Mills vacuum. Fermion fields in this theory have to be regarded as external sources. In terms of the link variables $U_{\mu}(x)$, for fermions in a representation $\R$ the Dirac operator is
\begin{eqnarray}
D_{xy} = 
m \delta_{xy} + \frac12 \sum_{\mu} \eta_\mu(x) \left\{ R[U_\mu(x)] \delta_{x+\hat{\mu},y} - R[U_\mu(x-\hat{\mu})]^\dagger \delta_{x-\hat{\mu},y} \right\} 
= m \delta_{xy} + K_{xy} \comma
\end{eqnarray}
$m$ being the quark mass in lattice units and the $\eta$'s the Kawamoto-Smit phases. Calling $\psi$ the spinor field, in terms of the Dirac operator, the chiral condensate is given by
\begin{eqnarray}
\langle \bar{\psi} \psi \rangle = \frac{1}{V} \langle \Tr D^{-1} \rangle = \frac{1}{Z V} \int \left( {\cal D} U \right) \left( \sum_{x} \tr D^{-1}_{x,x} \right) e^{- S} \ ,
\end{eqnarray}
where $V = \prod_i L_i$ is the lattice volume ($L_i$ being the size of the lattice in the $i$-th direction), $D^{-1}$ is the inverse staggered Dirac operator, $\Tr$ is the trace over both space and colour indices, and $\tr$ is the trace over colour indices.\\
In our specific case, orientifold planar equivalence states that for $N = \infty$ the condensate of $N_f$ quenched flavours of Dirac fermions of mass $m$ in the symmetric (or equivalently in the antisymmetric) representation of the gauge group is identical to the condensate of $N_f$ quenched Majorana flavours in the adjoint representation. In terms of Dirac flavours, this statement translates into
\begin{eqnarray}
\lim_{N \rightarrow \infty} \frac{1}{V N^2} \langle \Tr (m-K_{\Sym/\AS})^{-1} \rangle = 
\lim_{N \rightarrow \infty} \frac{1}{2 V N^2} \langle \Tr (m-K_{\Adj})^{-1} \rangle \ ,
\end{eqnarray}
where the factor of 1/2 on the r.h.s. is due to the replacement of Majorana flavours with Dirac flavours for the adjoint.\\ 
We can expand the quenched condensate for a generic representation $\R$ in powers of $1/N$:
\begin{flalign}\nonumber
\frac{1}{V N^2} \langle \Tr (m-K_{\R})^{-1} \rangle & = \frac{1}{V N^2} \sum_{n=0}^\infty \frac{1}{m^{n+1}} \langle \Tr K_{R}^n \rangle\\
& = \frac{1}{V N^2} \sum_{\omega \in \mathcal{C}} \frac{c(\omega)}{m^{L(\omega)+1}} \langle \tr \R[U(\omega)] \rangle \ ,
\end{flalign}
where the sum over $\omega$ runs over all closed paths $\mathcal{C}$ on the lattice, $L(\omega)$ is the length of the path $\omega$, $\R[U(\omega)]$ the path ordered product of the link variables in the representation $\R$ along $\omega$  and $c(\omega)$ the product of the Kawamoto-Smit phases along the same path. Recalling that
\begin{eqnarray}
\tr \Adj[U] = | \tr U |^2 -1 \qquad \mbox{and} \qquad \tr \Sym/\AS[U] = \frac{ (\tr U)^2 \pm \tr (U^2)}{2} \comma
\end{eqnarray}
we can write the condensates in the symmetric/antisymmetric and in the adjoint representation respectively as
\begin{flalign}
\label{condensates}
& \frac{1}{V N^2} \langle \Tr (m-K_{\Sym/\AS})^{-1} \rangle = \frac{1}{2V} \sum_{\omega \in \mathcal{C}} \frac{c(\omega)}{m^{L(\omega)+1}} \frac{\langle [\tr U(\omega)]^2 \rangle \pm \langle \tr [U(\omega)^2] \rangle}
{ N^2 }  \ ;\\
& \frac{1}{2 V N^2} \langle \Tr (m-K_{\Adj})^{-1} \rangle = \frac{1}{2V} \sum_{\omega \in \mathcal{C}} \frac{c(\omega)}{m^{L(\omega)+1}} \frac{\langle |\tr U(\omega)|^2 \rangle - 1 }{ N^2 } \ .
\end{flalign}
The leading terms at large $N$ are $[\tr U(\omega)]^2$ for the $\Sym/\AS$ condensate and $|\tr U(\omega)|^2$, both of order $N^2$. Hence, at large $N$ we can approximate the two previous equations with
\begin{flalign}
& \frac{1}{V N^2} \langle \Tr (m-K_{\Sym/\AS})^{-1} \rangle = \frac{1}{2V} \sum_{\omega \in \mathcal{C}} \frac{c(\omega)}{m^{L(\omega)+1}} \frac{\langle [\tr U(\omega)]^2 \rangle}
{ N^2 } \comma \\
& \frac{1}{2 V N^2} \langle \Tr (m-K_{\Adj})^{-1} \rangle = \frac{1}{2V} \sum_{\omega \in \mathcal{C}} \frac{c(\omega)}{m^{L(\omega)+1}} \frac{\langle |\tr U(\omega)|^2 \rangle}{ N^2 } \ .
\end{flalign}
In the $N \to \infty$ limit the two condensates are identical if
\begin{eqnarray}
\langle [\tr U(\omega)]^2 \rangle = \langle |\tr U(\omega)|^2 \rangle \ ,
\end{eqnarray}
or equivalently, using large $N$ factorisation, if
\begin{eqnarray}
\langle \tr U(\omega) \rangle = \langle \left( \tr  U(\omega) \right)^{\star} \rangle \ .
\end{eqnarray}
Thus, orientifold planar equivalence is a direct consequence of the invariance of the Yang-Mills theory under charge conjugation. Although such invariance is not in contradiction with any known result, at this stage it should be regarded as an assumption.
\section{Numerical Simulations}
The analytic proof holds at fixed lattice spacing $a$. This enables us to check orientifold planar equivalence at a given value of $a$. We choose the lattice spacing by requiring that for each SU($N$) group we have studied ($N=2,3,4,6,8$) the value of $\beta$ is the one at which the system displays the finite temperature deconfinement transition when the compactified direction is five lattice spacing long. These values of $\beta$ have been determined in~\cite{Lucini:2003zr} and recently used for a calculation of the meson spectrum at large $N$ in~\cite{DelDebbio:2007wk}. The simulations were performed on a $14^4$ lattice closed with antiperiodic boundary conditions in time and periodic boundary conditions in space. The physical size of the lattice is roughly (2 fm)$^4$.\\
The code used in our simulation is an adaptation of the HiRep code~\cite{DelDebbio:2008zf}. The advantage of this code over more popular lattice QCD packages is that both $N$ and the representation of the group can be set via a flag at compiling time. Moreover, since the code uses multishift solvers, we could determine the condensate for a large number of masses with little or no overhead with respect to the lightest mass. We choose $m$ in the range $(0.012;\ 8.0)$; these values allow us to check the equivalence from the chiral to the high mass regime.\\
An inspection of Eq.~(\ref{condensates}) suggests that the leading term contains only even powers of $1/N$ and the subleading term only odd powers of $1/N$. This is in contrast with the large $N$ limit in a pure Yang-Mills theory, where only even powers of $1/N$ appear. Indeed the absence of the $1/N$ and of the $1/N^3$ corrections in the latter can explain the observed rapid approach to the limiting theory for a wide range of observables (see~\cite{mtlat2008} for a review). From this argument, we can already expect slower convergence towards the large $N$ limit for the condensates. Unlike the symmetric/antisymmetric condensate, the adjoint condensate can be expanded as a power series in $1/N^2$, hence it should converge more rapidly to its large $N$ limit. In order to separate the even from the odd contributions in $1/N$, it is convenient to use the following parameterisation of the condensates:
\begin{flalign}
& \frac{1}{N^2} \langle \bar{\psi} \psi \rangle_{\Sym/\AS} = f \left( m, \frac{1}{N^2} \right) \pm  \frac{1}{N} g \left( m, \frac{1}{N^2} \right)  \ ,\\
& \frac{1}{N^2} \langle \lambda \lambda \rangle_{\Adj} = \tilde{f} \left( m, \frac{1}{N^2} \right)  - \frac{1}{2N^2} \langle \bar{\psi} \psi \rangle_{free}  \ ,
\end{flalign}
where the gauge singlet condensate $\langle \bar{\psi} \psi \rangle_{free}$ can be determined by diagonalising numerically the Dirac operator of a free fermion. In this language, orientifold planar equivalence means
\begin{equation}
f \left( m, 0 \right) = \tilde{f} \left( m, 0 \right) \ ,
\end{equation}
or equivalently
\begin{equation}
\label{pleq}
f \left( m, \frac{1}{N^2} \right) = \tilde{f} \left( m, \frac{1}{N^2} \right) + \frac{a}{N^2} + \frac{b}{N^4} + {\cal O}\left(\frac{1}{N^6}\right) \ .
\end{equation}
\begin{figure}
\begin{center}
\includegraphics[width=.6\textwidth]{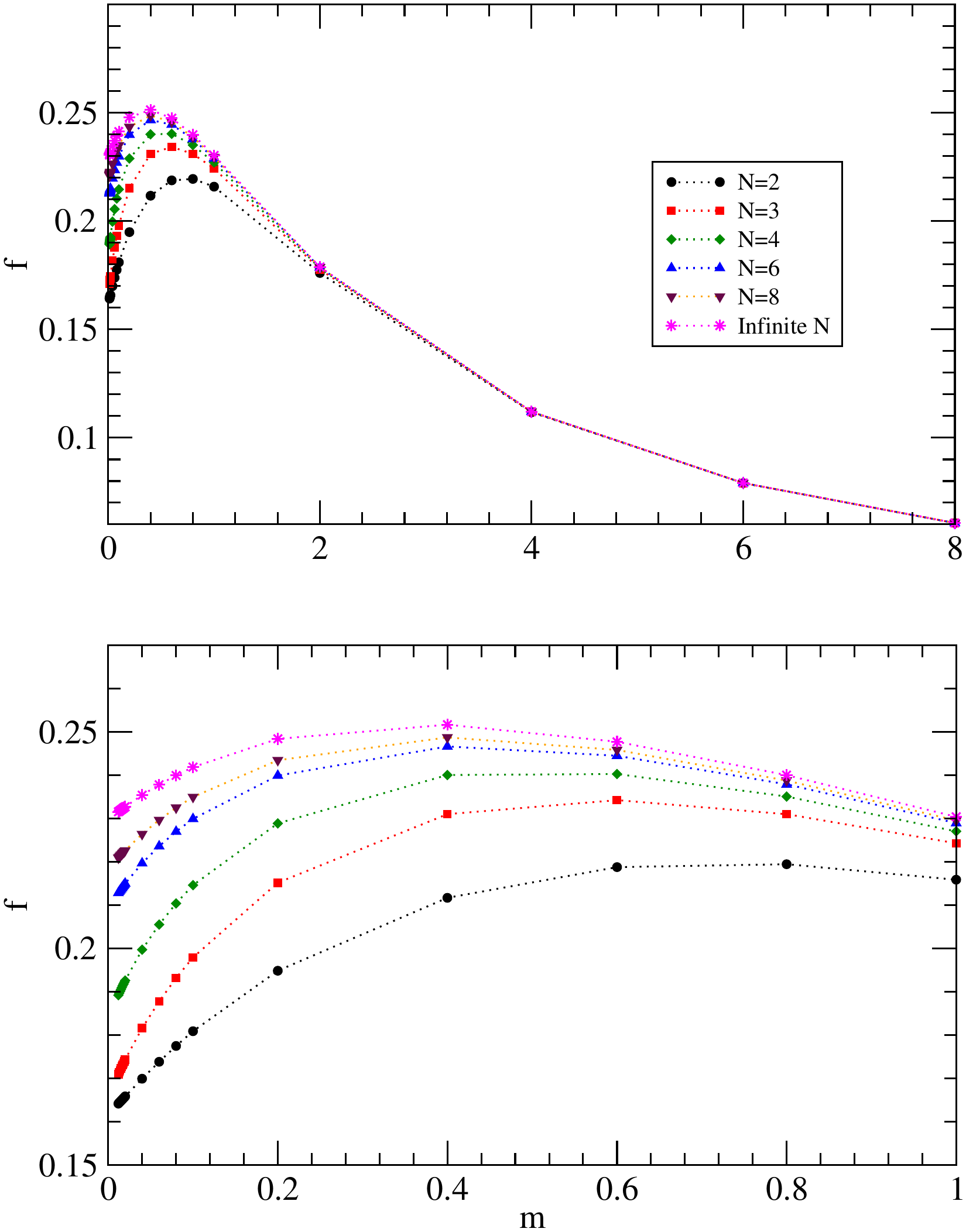}
\caption{Numerical results for the function $f$ at various $N$. The $N = \infty$ curve is also shown. The bottom figure is an enlargement of the top figure in the region $m \in (0; 1)$.}
\label{fig:f}
\end{center}
\end{figure}
\begin{figure}
\begin{center}
\includegraphics[width=.6\textwidth]{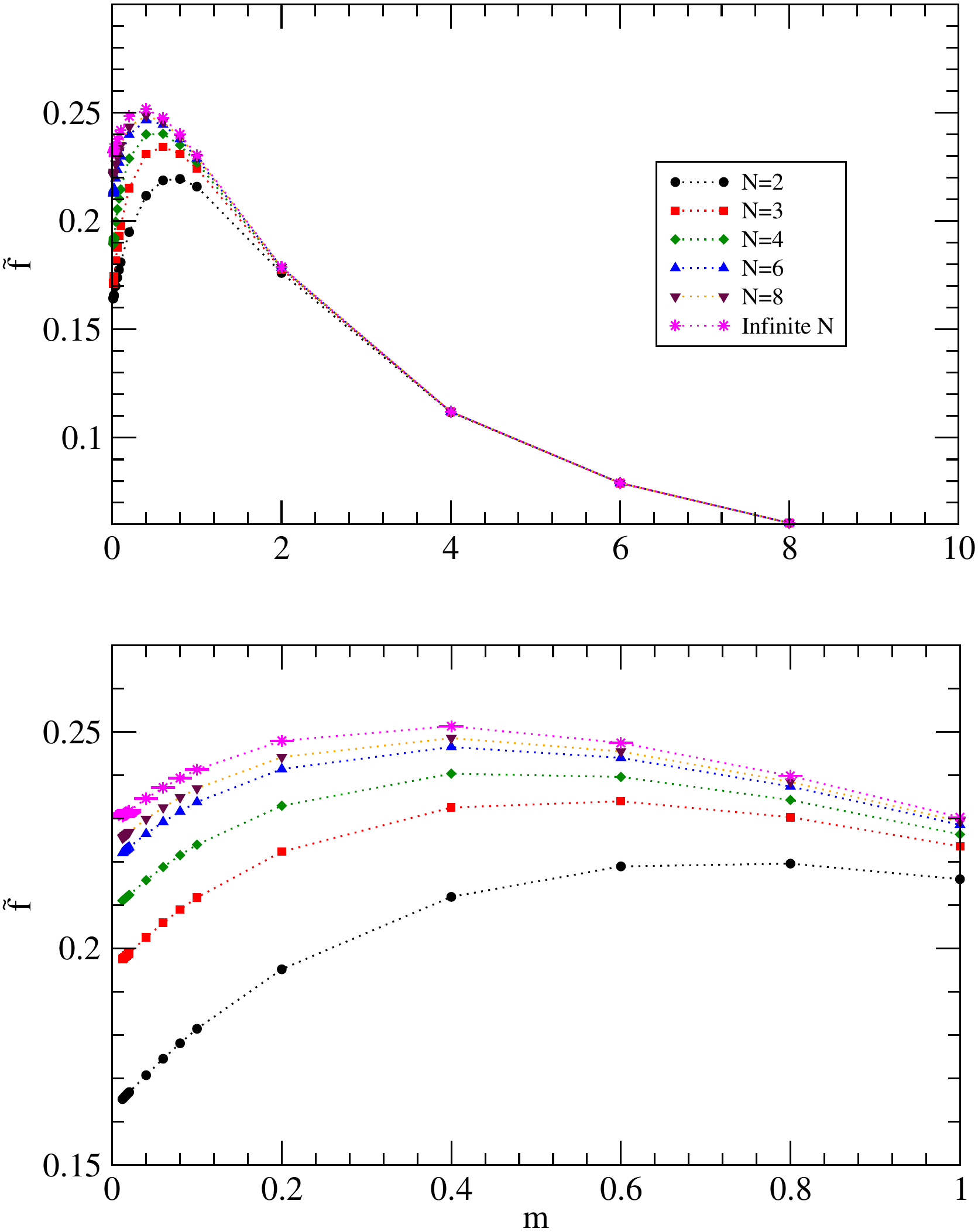}
\caption{Numerical results for the function $\tilde{f}$ at various $N$. The $N = \infty$ curve is also shown. The bottom figure is an enlargement of the top figure in the region $m \in (0; 1)$.}
\label{fig:ft}
\end{center}
\end{figure}
The functions $f$ and $g$ can be reconstructed from the condensates in the symmetric and antisymmetric representations; the condensate in the adjoint representation and the value of the condensate for the free Dirac field allow us to determine $\tilde{f}$. These functions can then be extrapolated to $N = \infty$ using the leading terms of the power expansion. In order to verify orientifold planar equivalence, we fitted our numerical data to Eq.~(\ref{pleq}). This ansatz worked very well, the reduced $\chi^2$ being of order one or less for any mass we have investigated. This proves non-perturbatively orientifold planar equivalence.\\
To address the question of the corrections at finite $N$, the functions $f$, $\tilde{f}$ and $g$ were determined. The large $N$ extrapolation worked well with only one leading correction for $\tilde{f}$, while for $f$ it was more effective to combine the results of $\tilde{f}$ with Eq.~(\ref{pleq}). Our numerical results for $f$ and $\tilde{f}$, including the extrapolations to infinite $N$, are displayed respectively in Fig.~\ref{fig:f}~and~\ref{fig:ft}.\\
It is instructive to use the extrapolated values for $f$, $\tilde{f}$ and $g$ to obtain an asymptotic prediction for the condensate. As an example, for $m = 0.012$ we get
\begin{align}
\nonumber
& \frac{ \langle \lambda \lambda \rangle_{\Adj} }{ N^2 } = 0.23050(22) - \frac{0.3134(72)}{N^2} + \dots \ ;\\
& \frac{ \langle \bar{\psi} \psi \rangle_{\Sym} }{ N^2 } = 0.23050(22) +\frac{0.4242(11)}{N} -\frac{0.612(43)}{N^2} +\frac{0.811(25)}{N^3} + \dots \ ;\\
\nonumber
& \frac{ \langle \bar{\psi} \psi \rangle_{\AS} }{ N^2 } = 0.23050(22) -\frac{0.4242(11)}{N} -\frac{0.612(43)}{N^2} -\frac{0.811(25)}{N^3} + \dots \ .
\end{align}
We can now compare those expressions with the corresponding numerical values at $N=3$. We find that while the analytical expressions for the adjoint and the symmetric agree respectively within 1\% and 4\% with the numerical results, the analytical formula for the condensate in the antisymmetric predicts a negative value at $N=3$, missing completely the numerical point. This is due to the large numerical cancellations among the leading terms that arise  at low $N$ in this representation, which increase the significance of the terms we have neglected. It is conceivable that a similar phenomenon happen also in the full theory.  

\section{Conclusions}
In this work, we have reported on the first numerical calculation involving (quenched) fermions in two-index representations at large $N$. In particular, we have proved numerically orientifold planar equivalence at fixed lattice spacing by computing the chiral condensates in the antisymmetric, symmetric and adjoint representations at various values of $N$ and extrapolating them to $N = \infty$. The lattice calculation enables us to compare the resulting asymptotic expansions with numerical data at finite $N$. We found that while for fermions in the adjoint and symmetric representations the leading corrections describe the data at $N \ge 3$ with an accuracy of a few percents or less, for fermions in the antisymmetric representation higher order corrections play a major role at $N = 3$. This suggests that in order to study numerically orientifold planar equivalence in the full theory, values of $N$ larger than those used for extrapolating observables in the Yang-Mills theory to $N = \infty$ might be needed.

\end{document}